\documentclass[prl,aps,twocolumn,showpacs]{revtex4}
\usepackage{epsfig}

\begin{document}

\title{GPS test of the local position invariance of Planck's constant}
\author{J. Kentosh}
\email{james.kentosh.336@my.csun.edu}
\author{M. Mohageg}
\email{makan.mohageg@gmail.com}

\affiliation{Department of Physics and Astronomy, California State University, Northridge, Northridge, California 91330-8268, USA}

\date{February 29, 2012}

\begin{abstract}
Publicly available clock correction data from the Global Positioning System was analyzed and used in combination with the results of terrestrial clock comparison experiments to confirm the local position invariance (LPI) of Planck's constant within the context of general relativity.  The results indicate that $h$ is invariant within a limit of $|\beta_h|<0.007$, where $\beta_h$ is a dimensionless parameter that represents the extent of LPI violation.
\end{abstract}

\pacs{06.20.Jr, 04.80.Cc, 06.30.Ft}

\maketitle

Many experiments and observations have tested possible variations of fundamental constants, particularly the fine-structure constant $\alpha$ \cite{uzan}.  Though most results have been null, studies of quasar absorption spectra suggest a small change in $\alpha$ over the past 8 billion years across the observable universe \cite{webb}.  If $\alpha$ were to vary as suggested by those results, that would imply a change in at least one of the parameters that comprise $\alpha = e^{2}/{{4} \pi \varepsilon_o \hbar c}$ \cite{bronn06}. In this letter, data from the Global Positioning System (GPS) is analyzed in combination with prior terrestrial clock comparison experiments \cite{tobar} to establish bounds on the invariance of Planck's constant $h$.

This analysis provides a test of `local position invariance' (LPI), which states that in local, freely falling frames the outcome of any nongravitational experiment is independent of where and when in the universe it is performed \cite{will}.  This principle is a variation of Einstein's equivalence principle relating gravitational fields to the accelerations of bodies in free space.  Nevertheless, as a macroscopic theory of gravity, general relativity tells us little about $h$.  Thus, experimental evidence of its invariance is necessary.

The Standard Model Extension (SME) has been developed partly to study possible variations in fundamental constants \cite{uzan, colladay, kostelecky}.  A less broad approach is used herein, based on the assumption that general relativity correctly describes macroscopic phenomena.  The observable effects of a variable $h$ are deduced from the invariance of the proper energy of atomic transitions. This approach provides a consistency check of general relativity, and our results may be relevant to the review of recent observations of super-luminal neutrinos \cite{CERN}.

The analysis proceeds in three steps:  First, we use GPS data to determine LPI limits for atomic clocks.  Then we incorporate the results of clock experiments to determine LPI limits for time dilation.  Finally we use both results to estimate the invariance of Planck's constant.

Using the standard convention (e.g., \cite{godone}) for expressing violations of LPI to first order in a gravitational field, the atomic transition frequency used in an atomic clock at $x$ and at some arbitrary reference point $O$ \--- such as a terrestrial laboratory \--- can be related as follows:\begin{equation}
f_{xo}/f_{o}=1+(1+\beta_{f})\Delta U/c^{2},
\end{equation}where $f_{o}$ is the frequency measured by an observer at point $O$ when the clock is at $O$, and $f_{xo}$ is the frequency when the clock is at $x$ as measured from the reference frame at $O$.  The potential difference is $\Delta U=U_{x}-U_{o}$, where $U_{i}=\Phi_{i}-v_{i}^{2}/2$, $\Phi_{i}$ is the gravitational potential energy per unit mass and $v_{i}$ is the clock's velocity.  The dimensionless parameter $\beta_{f}$ represents the extent of LPI violation for atomic clock frequencies.

The GPS uses atomic clocks on the ground and in orbit.  The GPS is operated on GPS time, a continuous timescale that can be related to Coordinated Universal Time (UTC).  To correct for clock drift, a clock correction for each satellite is broadcast as part of the navigation signal.  GPS satellite orbits have eccentricities generally less than 0.02 \cite{ashby96}.  At apogee, a satellite has a slower speed and higher gravitational potential, which cause its clock to run faster than at perigee and faster than a clock on the ground.  Corrections for this effect are  made in GPS, based on general relativity \cite{ashby96, kaplan}.

We analyzed GPS satellite data made available on the Internet by the International GNSS Service (IGS) \cite{dow}.  The IGS `Final' product in SP3-c format was used, in which GPS satellite positions and clock corrections are published for every 15 minutes of GPS time.  Positions are precise to 1 mm and clock corrections are precise to 1 ps \cite{hilla}.  A single SP3 file includes position and clock correction data for 32 GPS satellites for one GPS day.  The information in the files is based on data collected by IGS from the 32 satellites, 12 GPS control stations, and approximately 350 ground stations \cite{elrabbany}.

Of the 32 GPS satellites, seven were chosen that had more eccentric orbits and the most stable clocks:  2, 11, 18, 21, 26, 28, and 32.  Those satellites had the lowest ratios of the standard deviation of their changes in clock corrections to the theoretical range in clock rates for that satellite.  Satellite eccentricities ranged from 0.01 to 0.02.

The clock corrections published by IGS do not include the general relativistic corrections for the eccentricities of the GPS satellites.  It is left to the user to calculate and add those effects \cite{spec}.  Since we did not have to subtract it out, the missing relativistic component to the clock corrections simplified the analysis.

\begin{figure}[ht]
\center{
\epsfig{file=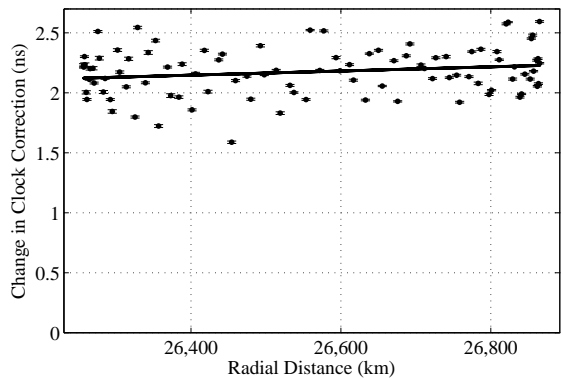, width=8.5cm, angle=0}}
\caption{\label{fitm} Changes in IGS-published clock corrections every 15 minutes for Satellite No. 18 on 6/10/10, plotted against distance from the earth's center.  The uncertainty in the ordinate is $\sigma$, as published in the SP3 files.  The least-squares linear fit has a small slope $m$.}
\end{figure}

To find $\beta_f$ the changes in the IGS-published clock corrections of the GPS satellites over each 15 minute interval were analyzed.  One year of such data was reduced, spanning the time period from April, 2010 through May, 2011, including extra days used for averaging.  For each satellite, 95 changes in its clock correction during each day were plotted as a function of its distance from the earth's center.  (The 96th spanned two SP3 files and was omitted for simplicity.)  A representative plot is shown in Fig. \ref{fitm}.

Since the relativistic eccentricity effect is not included in the published clock corrections, changes to those corrections should not vary with distance from the earth's center.  Standard theory predicts the slope $m$ of the linear fit in Fig. \ref{fitm} to be zero.  The small slope evident in Fig. \ref{fitm}, extracted by a linear least-squares regression fit to the data, might be explained by a non-zero $\beta_{f}$.  A non-zero slope could also be caused by any number of effects on satellite clock corrections, including random errors and atmospheric effects.

To estimate conservatively large bounds for $\beta_{f}$, all nonzero slopes were attributed to $\beta_{f}$.  The slope $m$ in Fig. \ref{fitm} is related to a possible non-zero value of $\beta_{f}$ by \begin{equation}
\beta_{f}=m(r_{a}-r_{p})/(2\Delta T_{max}),
\end{equation}where $r_{a}$ is the apogee radius, $r_{p}$ is the perigee radius and $\Delta T_{max}$ is the maximum theoretical clock correction of a satellite at apogee, given by the following, adapted from \cite{ashby03}:\begin{equation}
\Delta T_{max} =T_{ep}\left(\Phi_{a}-v_{a}^2/2-\Phi_{\text{GPS}}+v_{\text{GPS}}^{2}/2\right)/c^{2}.
\end{equation} In this equation, $\Phi_{a}$ and $v_{a}$ are the gravitational potential per unit mass and velocity of a satellite at apogee, respectively;  and $\Phi_{\text{GPS}}$ and $v_{\text{GPS}}$ are the same for a satellite in an ideal, circular GPS orbit corresponding to a radius of 26,561.75 km, representing GPS time.  (GPS satellites' clocks are adjusted to run more slowly pre-launch so that when they reach orbit, they run at GPS time.)  $T_{ep}$ is the `epoch time,' which is the 15 minute interval between clock corrections.

The coordinates in the SP3 files are published in an earth-fixed reference frame. A satellite's velocity relative to the earth's non-rotating frame was interpolated from its instantaneous radial distance, using Newtonian gravity. For an ideal, circular GPS orbit, this method reproduces the standard velocity $v_{\text{GPS}}$ of 3,873.83 m/s, used as a baseline in Eq. (3).

The mean offset of the changes in clock corrections from zero in Fig. \ref{fitm} indicates how much faster or slower a satellite's clock is advancing than GPS time.  That has no bearing on the current analysis.  Of interest is how the changes in clock corrections might vary with radial distance from the earth's center.

\begin{figure}[ht]
\center{
\epsfig{file=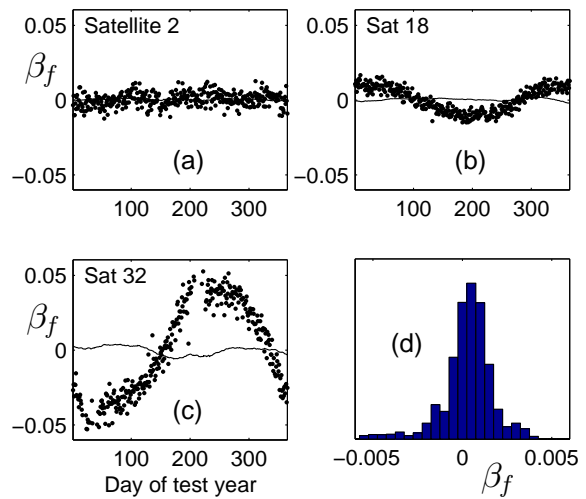, width=8.5cm, angle=0}}
\caption{\label{betaf} Daily and filtered values of $\beta_{f}$ for one year for (a) Sat 2, (b) Sat 18, (c) Sat 32.  Sat 2 had the least annual oscillation; Sat 32 had the greatest range in $\beta_f$.  (d) Combined probability distribution of filtered $\beta_f$ for 7 satellites for one year.}
\end{figure}

The best fit slope and the corresponding daily values for $\beta_{f}$ were calculated for the seven satellites, for each day in the test period, with 2,749 plots generated.  (The clock data was partially missing or corrupted on some days.)  Results for three representative satellites are shown in Fig. \ref{betaf}(a-c).

The plots of $\beta_{f}$ exhibit significant daily variations, and an annual oscillation that varied among satellites.  The daily variations were consistent in magnitude among the satellites.  The annual oscillations are likely caused by atmospheric effects.  The maxima of these oscillations roughly coincide with a semi-major axis aligned with Eurasia and the Pacific Ocean, an orientation that shifts 360 degrees over a year.

Time-series and Fourier analysis of the data was used to develop appropriate digital filters to remove yearly, bi-monthly and daily variations driven by nongravitational effects. To remove the annual oscillation, a least-squares fit was computed using $y=A+B\cos(2\pi t/1yr+\phi)$, where $A$, $B$ and $\phi$ are fit parameters.  The $B\cos(2\pi t/1yr+\phi)$ term was subtracted from the daily values.  Based on the frequency distribution in the Fourier analysis, we then used a 51-day moving average to filter out short-term fluctuations in $\beta_{f}$.  The filtered data for three satellites (2, 11, 32) exhibited a sawtooth-shaped oscillation with a 120-day period, of unknown origin.  That oscillation was filtered out for those three satellites in the same way as the annual oscillations.  The limits on $\beta_f$ were most sensitive to filtering out the annual oscillations, but less sensitive to the duration of the rolling average or to removing the 120-day oscillations. The residual daily values of $\beta_f$ for three satellites are shown with solid lines in Fig. \ref{betaf}(a-c).

The residual values of $\beta_f$ from seven satellites exhibit the bell-shaped distribution shown in Fig. \ref{betaf}(d).  A small residual mean of $\beta_{f}=0.0003$ was present.  We cannot rule out GPS systemic errors as its source.  Based on a 95\% confidence level, the following limits were found: $-0.0027<\beta_{f}<0.0033$.  The results for the seven satellites were weighted equally.  Finding these limits on $\beta_{f}$ is the first step in setting limits on the invariance of Planck's constant.

The well-behaved annual oscillations in Fig. \ref{betaf} demonstrate some statistical significance for the linear fits for $m$.  However, values of $R^{2}$ for the fits ranged from $10^{-5}$ to $0.6$ with a mean of $0.04$, suggesting a low to moderate statistical significance for $m$ \cite{taylor}.

To study this effect, a statistical model was developed in which the standard deviations of the GPS clock corrections were used to create a randomly-generated data set to mimic the GPS data.  Ninety-five randomly generated points were used to generate fits for $m$, repeated 2,555 times and then filtered to model our analysis.  The results indicate that about 2/3 of the limits on $\beta_f$ arise from random variations in the GPS clock corrections.  However, some GPS satellites exhibit more consistent clock corrections and the bounds could be reduced by analyzing only those satellites.

If $h$ were to vary, general relativistic time could differ from atomic time.  Possible LPI violations of gravitational time dilation can be expressed to first order as\begin{equation}
dt_{x}/dt_o=1+(1+\beta_t)\Delta U/c^2,
\end{equation}where $dt_{x}/dt_{o}$ is the ratio of the time rates at $x$ and $O$, and $\beta_{t}$ is a dimensionless parameter that represents the extent of LPI violation for time dilation.  If $\beta_{t}=0$, the formula reduces to general relativity.

Cryogenic optical resonator clocks are another type of clock used in tests of fundamental constants.  Unlike atomic clocks, the rates of these clocks depend on the speed of light rather than Planck's constant.  An invariant $c$ means that the locally-measured (proper) rates of suitably configured resonator clocks will not vary with position, even if $h$ varies.  (Some resonator clocks are locked to an atomic frequency and would depend on $h$ \cite{maleki}.)  Therefore, a resonator clock measures proper time and can be used to determine $\beta_{t}$.

Comparisons between atomic clocks (hydrogen masers) and cryogenic sapphire oscillators (CSO) have been performed at the Paris Observatory \cite{tobar} and other laboratories (e.g., \cite{turn}).  Tobar et al. summarize that work and show that little measurable difference is observed as the earth moves in its elliptical orbit within the gravitational potential of the sun, within a limit of $\beta_{\text{H maser}}-\beta_{\text{CSO}}=-2.7(1.4)\times 10^{-4}$ for annual variations \cite{tobar}.  The relative rates of atomic and resonator clocks differ little with gravitational potential.

Once a GPS satellite reaches orbit, the clock experiments \cite{tobar} indicate that a resonator clock and an atomic clock on the satellite would advance at the same relative rates.  Therefore, the variations in clock corrections should apply equally to resonator clocks, and can be used to determine a limit on $\beta_{t}$.  In other words, within a GPS orbit we can treat GPS atomic clocks as a surrogate for resonator clocks, so that $\beta_t \approx \beta_f$.

In the clock comparison experiments, $\beta_{\text{H maser}}-\beta_{\text{CSO}}$ corresponds to $\beta_{f}-\beta_{t}$, and their limits were added to the limits on $\beta_{f}$ to determine limits for $\beta_{t}$.  Combining the maximum uncertainties, the limits on LPI violation for time dilation were found to be $-0.0031<\beta_{t}<0.0037$.

To prove that $h$ is invariant in the context of SME theory is probably not possible with existent experimental evidence.  However, in general relativity the locally-measured (proper) rest energy of a particle is independent of gravitational potential.  Based on arguments similar to Nordtvedt's \cite{nordtvedt}, the proper energy emitted by a specific atomic transition should also be invariant.  For example, the locally measured energy emitted by the transition between the two hyperfine levels of the ground state of $\text{Cs}^{133}$ \--- the isotope used in atomic clocks \--- should be the same at all points and time in the universe.

With a locally-invariant emission energy $E_o$, the proper frequency of emission $f_{x}$ of a given atomic transition measured at any elevation $x$ would then be given by \begin{equation}
f_x=E_o/h_x,
\end{equation}where $h_{x}$ is the locally measured value of $h$ at $x$.  If $h$ satisfies LPI, then $h_{x}$ would be constant and the proper frequency of an atomic clock would also be invariant.  However, if $h_{x}$ varies with position then the proper frequency of an atomic clock would also vary.

LPI violations for $h$ can be written as\begin{equation}
h_x/h_o=1+\beta_h\Delta U/c^2,
\end{equation}where $h_{o}$ is the locally measured value of $h$ at reference point $O$, $h_{x}$ is its locally measured value at $x$, and $\beta_{h}$ is the parameter for Planck's constant.

Let $f_{o}$ be the proper frequency of an atomic clock when it is at $O$.  Since $E_{o}=h_{o}f_{o}$, it follows from Eq. (5) that at $x$,\begin{equation}
f_x=h_{o}f_{o}/h_x.
\end{equation}

Due to time dilation, the frequency $f_{xo}$ of the clock at $x$ measured by an observer at $O$ will be \begin{equation}
f_{xo}=(dt_x/dt_o)(h_{o}f_o/h_x).
\end{equation}

A formula relating the three $\beta'$s to first order can be found by substituting Eqs. (1), (4) and (6) into (8), which yields $\beta_{h}=\beta_{t}-\beta_{f}+O(\mu^{2})$, where $\mu=\Delta U/c^{2}$.  Second order relativistic effects in the earth's field are negligible, so higher order terms can be ignored, leaving\begin{equation}
\beta_h \approx \beta_t-\beta_f.
\end{equation}

The traditional interpretation of general relativity predicts that $\beta_{t}=\beta_{f}\equiv 0$, yielding $\beta_{h}=0$, which would signify that $h$ is invariant.


The estimated limits on $\beta_{t}$ and $\beta_{f}$ were used in Eq. (9) to determine LPI limits for Planck's constant.  Combining the limits on those two parameters and accounting for their signs, our final results, accurate to one significant digit, are summarized in Table I.

\begin{table}[ht]
\caption{Estimated limits on LPI violation, based on this study unless otherwise noted.}
\centering
\begin{tabular}{l c c}
\hline\hline
Type of LPI & Limits found & Basis \\[0.5ex]
\hline
Atomic clocks & $-0.0027<\beta_f<0.0033$ & GPS data \\
Clock comparisons & $\beta_f-\beta_t=-2.7(1.4)\times 10^{-4}$ & Ref. [4] \\
Time dilation & $-0.0031<\beta_t<0.0037$ & GPS \& [4] \\
Planck's constant & $|\beta_h|<0.007$ & Eq. (9) \\
\hline\hline
\end{tabular}
\end{table}

To illustrate the effects of the filtering, we obtain $|\beta_h|<0.01$ if we use a 1-year filter but substitute an 11-day moving average of $\beta_f$ (smaller than the 28-day lunar cycle) for the 51-day average and omit the 120-day filters.

In conclusion, general relativity requires the local invariance of the speed of light and emission energy.  Within those constraints we find, based on experimental evidence from the GPS and clock comparison experiments, that $h$ satisfies local position invariance to within a limit of 0.007.  The data also supports the gravitational time dilation of general relativity for geometrodynamic clocks within a limit of 0.004.

The work reported in this letter was carried out at California State University, Northridge.


\end{document}